\documentclass{article} 
 
\usepackage{amsmath,amssymb,amsthm,latexsym} 
 
\usepackage{stmaryrd,wasysym,upgreek,mathrsfs,dsfont} 
\usepackage[english]{babel} 
\usepackage{graphicx,color}

\usepackage[pdftex]{hyperref}

\newtheorem{theorem}{Theorem}[section]
\newcommand{\beq}{\begin{equation}}
\newcommand{\eeq}{\end{equation}}
\newcommand{\beqa}{\begin{eqnarray}}
\newcommand{\eeqa}{\end{eqnarray}}

\begin{document} 
 \title{Existence of Asymptotic Expansions in Noncommutative Quantum Field Theories} 
\author{C.A. Linhares$^{(1)}$, A.P.C. Malbouisson$^{(2)}$ and 
I. Roditi$^{(2)}$\\ 
$^{(1)}$Instituto de F\'{\i}sica, Universidade do Estado do Rio de Janeiro, \\
Rua S\~{a}o Francisco Xavier, 524, 20559-900 Rio de Janeiro, RJ, Brazil   \\ 
$^{(2)}$Centro Brasileiro de Pesquisas F\'{\i}sicas,\\
Rua Dr.~Xavier Sigaud, 150, \\ 
22290-180 Rio de Janeiro, RJ, Brazil}
 
\maketitle 

\begin{abstract}
Starting from the complete Mellin representation of Feynman amplitudes for
noncommutative vulcanized scalar quantum field theory, introduced in a previous publication,
we generalize to this theory the study of asymptotic behaviours under
scaling of arbitrary subsets of external invariants of any Feynman
amplitude. This is accomplished for both convergent and renormalized amplitudes.
\end{abstract}

\section{Introduction}

The possibility of studying both the ultraviolet and infrared behaviours of
Feynman amplitudes in quantum field theories, obtained directly without the
need of first calculating explicitly the complete expressions for them, is a
subject that is still finding new applications. In the present paper, we
seek such asymptotic expansions for noncommutative theories  known in the literature as of the `vulcanized' type, that is, those which incorporate suitable modifications
in order to avoid the occurrence of the ultraviolet--infrared divergence
mixing, and thus become renormalizable \cite
{grosse1,grosse2,grosse3,rivasseau1,rivasseau2,vignes}.

The direct approach to asymptotic behaviours was formulated for commuting
theories in the 1970's in the papers \cite{bergere1,bergere2,bergere3}
within the Bogoliubov--Parasiuk--Hepp--Zimmermann renormalization scheme. It
is based on the Feynman--Schwinger parametric representation of amplitudes,
expressed in terms of Symanzik polynomials in the Schwinger parameters \cite
{nakanishi,itzykson}.

However, in vulcanized noncommutative theories, propagators are based on the
Mehler kernel, instead of the heat kernel of commutative theories. This
leads to propagators that are quadratic in the position space, so that the
noncommutative parametric representation involves integration over position
and momentum variables, which can be performed. It results that one obtains
hyperbolic polynomials in the Schwinger parameters, not just the Symanzik
polynomials of the commutative case \cite{rivasseau3,rivasseau4}. See also
the reviews \cite{rivasseau5,rivasseau5a,rivasseau6}.

In \cite{bergere1,bergere2,bergere4}, the Mellin transform technique was
applied in order to prove theorems implying the existence of asymptotic
expansions of the amplitudes and in \cite{bergere5} the concept of `FINE'
polynomials was introduced, that is, those having the property of being
factorizable in each Hepp sector \cite{hepp} of the variables (a complete
ordering of the Schwinger parameters). Under scaling by a parameter $\lambda 
$ of (at least a few of) external invariants associated to a diagram, the
Mellin transform with respect to this scaling parameter leads, as $\lambda $
is taken to infinity, to an asymptotic series in powers of $\lambda $ and
powers of logarithms of $\lambda $. This was possible because for amplitudes
having the FINE\ property the Mellin transform may be `desingularized',
which means that the integrand of the inverse Mellin transform, which gives
back the Feynman amplitude as a function of $\lambda $, has a meromorphic
structure, so that the residues of its various poles generate the asymptotic
expansion. However, this is not the case under arbitrary scaling, as the
FINE\ property simply does not occur in many diagrams.

For those non-FINE\ diagrams, it was introduced in \cite{bergere5} the
so-called `multiple Mellin' representation, which consists in splitting the
Symanzik polynomials in a certain number of pieces, each one of which having
the FINE\ property. Then, after scaling by the parameter $\lambda $, an
asymptotic expansion can be obtained as a sum over all Hepp sectors. This is
always possible to do if one adopts, as in \cite
{decalan1,decalan2,decalan3}, the extreme point of view of splitting the
Symanzik polynomials in all its monomials, which leads to the so-called  `complete
Mellin' (CM) representation. The CM representation provides a general
framework to the study of asymptotic expansions of Feynman amplitudes.
Moreover, the integrations over the Schwinger parameters can be explicitly
performed without any division of the integral into Hepp sectors, and we are
left with the pure geometrical study of convex polyhedra in the Mellin
variables \cite{decalan1}. Also, the CM representation allows a unified
treatment of the asymptotic behaviour of both ultraviolet convergent and
divergent amplitudes. This happens because, as shown in \cite
{decalan1,decalan2}, the renormalization procedure does not alter the
algebraic structure of integrands in the CM\ representation. It only changes
the set of relevant integration domains in the Mellin variables. The method
allows the study of dimensional regularization \cite{decalan2,decalan3} and
of the infrared behaviour of amplitudes relevant to critical phenomena \cite
{malbouisson}. With the CM representation one is also able to prove the
existence of asymptotic expansions for most useful commutative field
theories, including gauge theories in an arbitrary gauge \cite{linhares}.

In what regards noncommutative field theories, one expects that an
adaptation of the general results of all these references could be
developed. In fact, recently \cite{rivasseau7}, the CM representation has
been extended to the `vulcanized' noncommutative $\phi ^{\star 4}$ massless
theory and a proof of dimensional meromorphy of its Feynman amplitudes has
been presented. Our choice of a massless theory is due to the fact that the
CM representation becomes less explicit and less appealing in the massive
model. In any case, masses are not essential for vulcanized noncommutative
field theories which have no `infrared divergences' and only
`half-a-direction' for their renormalization group. Based on Ref.~\cite
{rivasseau7}, in the present paper we intend to show that asymptotic
expansions exist for this noncommutative theory, in a similar way as the
analogous result for the respective commutative theory. We also study
explicitly the case of divergent noncommutative amplitudes in the CM
representation, by adapting to this context the renormalization procedure of
subtraction of suitably truncated Taylor expansions of amplitude integrand
functions along the lines of Refs. \cite{bergere1,decalan1,decalan2,gurau}.
We find that the renormalization procedure in the CM\ representation, as
already mentioned for commutative theories, also does not alter the
algebraic structure of integrands for the noncommutative Feynman amplitudes,
only the set of relevant integration domains in the Mellin variables
changes. This allows to transpose to divergent Feynman integrals the
machinery used in the convergent case and prove the existence of asymptotic
expansions for renormalized amplitudes. 

The paper is organized as follows. In section 2, we very briefly recall the main 
features of the 
complete Mellin representation for commutative scalar theories. Next, in section 3,
we review the CM representation for the vulcanized $\phi^{\star 4}$ theory. In
sections 4 and 5 we present the generalizations to the noncommutative theory of the respective theorems on the existence of 
the asymptotic expansions for the convergent and renormalized amplitudes. In the last
section we summarize our conclusions.

\section{Complete Mellin representation in the commutative scalar case}

Let us first consider the simpler case of a Feynman amplitude in a
commutative massive scalar theory. The amplitude related to an arbitrary
diagram $G$, with $I$ internal lines, $V$ vertices, and $L$ loops, in $d$ 
spacetime dimensions, reads
\begin{equation}
\mathcal{A}_G=C_{G}\int_0^\infty \frac{\prod_{\ell =1}^Id\alpha _\ell }{\left(
4\pi \right) ^{dL/2}U^{d/2}(\alpha )}e^{-\sum_\ell \alpha _\ell m_\ell
^2}e^{-N(s_k;\alpha )/U(\alpha )},
\end{equation}
where $C_{G}$ is a constant, $U$ and $N$ are homogeneous polynomials in the $\alpha _\ell $
variables, known in the literature as the Symanzik polynomials, which are
written as 
\begin{equation}
U(\alpha )=\sum_j\prod_{\ell =1}^I\alpha _\ell ^{u_{\ell j}}\equiv
\sum_jU_j,\qquad N(\alpha )=\sum_ks_k\left( \prod_{\ell =1}^I\alpha _\ell
^{n_{\ell k}}\right) \equiv \sum_kN_k,
\end{equation}
where $j$ runs over the set of 1-trees and $k$ over the set of 2-trees of
the diagram $G$; $s_k$ are $O(d)$-invariants given by the square of the sum
of all external momenta at one of the components of the 2-tree $k$; also, 
\begin{equation}
u_{\ell j}=\left\{ 
\begin{array}{lll}
0 &  & \text{if the line }\ell \text{ belongs to the 1-tree }j \\ 
1 &  & \text{otherwise}
\end{array}
\right. 
\end{equation}
and 
\begin{equation}
n_{\ell k}=\left\{ 
\begin{array}{lll}
0 &  & \text{if the line }\ell \text{ belongs to the 2-tree }k \\ 
1 &  & \text{otherwise.}
\end{array}
\right. 
\end{equation}
The complete Mellin representation for $\mathcal{A}_G$, following the steps
shown in \cite{decalan1,decalan2,linhares} is given by 
\begin{equation}
\mathcal{A}_G(s_k,m_\ell ^2)=\int_\delta \frac{\prod_j\Gamma (-x_j)}{\Gamma
\left( -\sum_jx_j\right) }\prod_ks_k^{y_k}\Gamma \left( -y_k\right)
\prod_\ell \left( m_\ell ^2\right) ^{-\phi _\ell }\Gamma \left( \phi _\ell
\right) ,  \label{Melcom}
\end{equation}
where 
\begin{equation}
\phi _\ell =\sum_ju_{\ell j}x_j+\sum_kn_{\ell k}y_k+1.
\end{equation}
The symbol $\int_\delta $ means integration over the independent variables $%
\frac{\text{Im }x_j}{2\pi i}$, $\frac{\text{Im }y_k}{2\pi i}$ in the convex
domain $\delta $ defined by ($\sigma $ and $\tau $ standing respectively for
Re $x_j$ and Re $y_k$) 
\begin{equation}
\delta =\left\{ \sigma ,\tau \left| 
\begin{array}{l}
\sigma _j<0;\,\tau _k<0;\,\sum_jx_j+\sum_ky_k=-\frac d2; \\ 
\forall i,\text{ Re }\phi _i=\sum_ju_{ij}\sigma _j+\sum_kn_{ik}\sigma _k+1>0
\end{array}
\right. \right\} .
\end{equation}
This domain $\delta $ is nonempty as long as $d$ is positive and small
enough so that every subdiagram of $G$ has convergent power counting \cite
{decalan1}; hence in particular for the $\phi ^4$ theory it is always
nonempty for any diagram for $0<d<2$.

Let us denote collectively by $\zeta _\mu $ the arguments of the $\Gamma $%
-functions: $-x_j$, $-y_k$, $\phi _\ell $. Also we call collectively $t_\mu $
the set of invariants $s_k$ and the squared masses $m_\ell ^2$. A general
asymptotic regime is then defined as the scaling $t_\mu \rightarrow \lambda
^{b_\mu }t_\mu $, in such a way that the amplitude (\ref{Melcom}) becomes a
function of $\lambda $ written in the convenient form \cite{decalan1}
\begin{equation}
\mathcal{A}_G(\lambda )=\int_\delta \lambda ^\zeta \prod_\mu t_\mu ^{-{\zeta 
}_\mu }\Gamma (\zeta _\mu )\frac 1{\Gamma \left(
-\sum_jx_j\right) },  \label{lambfunc}
\end{equation}
with $\zeta =\sum_\mu b_\mu \zeta _\mu $. This representation can be
extended to complex values of $d$. For instance, for a massive $\phi ^4$
diagram, it is analytic in $d$ for Re $d<2$ and meromorphic in $d$ in the
whole complex plane with singularities at rational values; furthermore, its
dimensional analytic continuation has the same unchanged CM integrand but
translated integration contours. Also, it is valid without change in the
form of the integrand for \textit{renormalized} amplitudes \cite
{decalan1,decalan2}. Using the meromorphic properties of the integrand of 
eq.~(\ref{lambfunc}), an asymptotic expansion in powers of $\lambda $ and 
powers of logarithms of $\lambda $ is obtained for $\mathcal{A}_G(\lambda )$ 
in Ref~\cite{decalan1}. 

\section{Complete Mellin representation for noncommutative scalar theories}

In order to establish notation, we review in this section the results of 
\cite{rivasseau7}, which we take as the starting point of the study of
asymptotic behaviours and renormalization, to be developed in the following
sections, and which constitutes the main subject of the present paper.

According to the analysis exposed in \cite{rivasseau3}, the amplitude
related to a ribbon diagram $G$ with $L$ internal lines, by choosing a
particular root vertex $\bar{V}$, has a parametric representation in terms
of the variable $t_\ell =\tanh \alpha _\ell /2$, where $\alpha _\ell $ are
the former Schwinger parameters as 
\begin{equation}
\mathcal{A}_G(\{x_e\},p_{\bar{V}})=\mathcal{K}_G\int_0^1\prod_\ell dt_\ell
\,(1-t_\ell ^2)^{d/2-1}\int dx\,dp\,\exp \left[ -\frac \Omega 2X\mathcal{G}%
X^t\right] ,
\end{equation}
where $\mathcal{K}_G$ is a constant, $d$ is the spacetime dimension, $\Omega $ is the Grosse--Wulkenhaar
vulcanization coefficient, $X$ summarizes all positions and hypermomenta, and 
$\mathcal{G}$ is a certain quadratic form. Calling $x_e$ and $p_{\bar{V}}$
the external variables and $x_i$, $p_i$ the internal ones, we decompose $%
\mathcal{G}$ into an internal quadratic form $Q$, an external one $M$ and a
coupling part $P$, so that 
\begin{equation}
X=\left( 
\begin{array}{llll}
x_e & p_{\bar{V}} & x_i & p_i
\end{array}
\right) ,\qquad \mathcal{G}=\left( 
\begin{array}{ll}
M & P \\ 
P^t & Q
\end{array}
\right) .
\end{equation}
Performing the Gaussian integration over all internal variables, one gets
the noncommutative parametric representation given by 
\begin{equation}
\mathcal{A}_G(\{x_e\},p_{\bar{V}})=\mathcal{K}_G\int_0^1\prod_\ell dt_\ell
\,(1-t_\ell ^2)^{d/2-1}\frac{e^{-HV_{G,\bar{V}}(t,x_e,p_{\bar{V}})/HU_{G,%
\bar{V}}(t)}}{\left[ HU_{G,\bar{V}}(t)\right] ^{d/2}},  \label{param}
\end{equation}
where new polynomials, in the $t_\ell $ variables ($\ell =1,\ldots ,L$), $%
HU_{G,\bar{V}}$ and $HV_{G,\bar{V}}$, have been introduced, which are the
analogs of the Symanzik polynomials $U$ and $N$ of the commutative case. It
has been shown in \cite{rivasseau3} that for the Grosse--Wulkenhaar $\phi
^{\star 4}$ model we have 
\begin{eqnarray}
HU_{G,\bar{V}} &=&\sum_{K_U=I\cup J;\,n+|K_U|\text{ odd}%
}s^{2g-k_{K_U}}n_{K_U}^2\prod_{\ell \notin I}t_\ell \prod_{\ell ^{\prime
}\in J}t_{\ell ^{\prime }}  \nonumber \\
&=&\sum_{K_U}a_{K_U}\prod_\ell t_\ell ^{u_{\ell K_U}}\equiv
\sum_{K_U}HU_{K_U},
\end{eqnarray}
where $I$ is a subset of the first $L$ indices, with $|I|$ elements, and $J$
a subset of the next $L$ indices, with $|J|$ elements; $s=1/4\Theta \Omega $
is a constant containing the noncommutative parameter $\Theta $ and the
vulcanization coefficient $\Omega $; $g$ is the genus of the diagram, $%
a_{K_U}=s^{2g-k_{K_U}}n_{K_U}^2$, $k_{K_U}=|K_U|-L-F-1$, $F$ being the
number of faces of the diagram; $n_{K_U}=\mathrm{Pf}\left( B_{\hat{K}%
_U}\right) $, where $B$ is the antisymmetric part of the quadratic form $Q$
restricted by omitting hypermomenta, so $n_{K_U}$ is the Pfaffian of the
antisymmetric matrix obtained from $B$ by deleting the lines and columns in
the set $K_U=I\cup J$; finally, 
\begin{equation}
u_{\ell K_U}=\left\{ 
\begin{array}{lll}
0 &  & \text{if }\ell \in I\text{ and }\ell \notin J \\ 
1 &  & \text{if }\ell \notin I\text{ and }\ell \notin J\text{ } \\ 
2 &  & \text{if }\ell \notin I\text{ and }\ell \in J.
\end{array}
\right. 
\end{equation}

The second polynomial $HV$ has both a real part $HV^R$ and an imaginary part 
$HV^I$. We need to introduce beyond $I$ and $J$ as above a particular line $%
\tau \notin I$ which is the analog of a 2-tree cut. Then it is shown in \cite
{rivasseau3} that 
\begin{eqnarray}
HV_{G,\bar{V}}^R &=&\sum_{K_V=I\cup J}\prod_{\ell \notin I}t_\ell
\prod_{\ell ^{\prime }\in J}t_{\ell ^{\prime }}\left[
\sum_{e_1}x_{e_1}\sum_{\tau \notin K_V}P_{e_1\tau }\epsilon _{K_V\tau }\text{%
Pf}\left( B_{\hat{K}_V\hat{\tau}}\right) \right] ^2  \nonumber \\
&=&\sum_{K_V}s_{K_V}^R\left( \prod_{\ell =1}^Lt_\ell ^{v_{\ell K_V}}\right)
\equiv \sum_{K_V}HV_{K_V}^R,
\end{eqnarray}
where 
\begin{equation}
s_{K_V}^R=\left[ \sum_ex_e\sum_{\tau \notin K_V}P_{e\tau }\epsilon _{K_V\tau
}\text{Pf}\left( B_{\hat{K}_V\hat{\tau}}\right) \right] ^2
\end{equation}
and $v_{\ell K_V}$ is given by the same formula as $u_{\ell K_U}$. The
imaginary part involves pairs of lines $\tau $, $\tau ^{\prime }$ and
corresponding signatures \cite{rivasseau3, rivasseau4}: 
\begin{eqnarray}
HV_{G,\bar{V}}^I &=&\sum_{K_V=I\cup J}\prod_{\ell \notin I}t_\ell
\prod_{\ell ^{\prime }\in J}t_{\ell ^{\prime }}\epsilon _{K_V}\text{Pf}%
\left( B_{\hat{K}_V}\right)   \nonumber \\
&&\times \left[ \sum_{e_1,e_2}\left( \sum_{\tau ,\tau ^{\prime }}P_{e_1\tau
}\epsilon _{K_V\tau \tau ^{\prime }}\text{Pf}\left( B_{\hat{K}_V\hat{\tau}%
\tau ^{\prime }}\right) P_{e_2\tau ^{\prime }}\right) x_{e_1}\sigma
x_{e_2}\right]   \nonumber \\
&=&\sum_{K_V}s_{K_V}^I\left( \prod_{\ell =1}^Lt_\ell ^{v_{\ell K_V}}\right)
\equiv \sum_{K_V}HV_{K_V}^I,
\end{eqnarray}
where 
\begin{equation}
s_{K_V}^I=\epsilon _{K_V}\text{Pf}\left( B_{\hat{K}_V}\right) \left[
\sum_{e,e^{\prime }}\left( \sum_{\tau ,\tau ^{\prime }}P_{e\tau }\epsilon
_{K_V\tau \tau ^{\prime }}\text{Pf}\left( B_{\hat{K}_V\hat{\tau}\tau
^{\prime }}\right) P_{e^{\prime }\tau ^{\prime }}\right) x_e\sigma
x_{e^{\prime }}\right] ,
\end{equation}
where $\sigma =\left( 
\begin{array}{ll}
\sigma _2 & 0 \\ 
0 & \sigma _2
\end{array}
\right) $ and $\sigma _2$ is the second Pauli matrix.

The main differences of the noncommutative parametric
representation with respect to the commutative case are the presence of the
constants $a_{K_U}$ in $HU$ (which contains the noncommutative quantity $s=1/4\Theta \Omega $), the presence of the imaginary part $iHV^I$ in $HV$,
and the fact that the parameters $u_{\ell j}$ and $v_{\ell k}$ in the
formulas above can have also the value 2 (and not only 0 and 1).

In order to proceed, we now introduce the Mellin parameters. 
For the real part $HV^R$ of $HV$, we use the identity \cite{rivasseau7}
\begin{equation}
e^{-HV_{K_V}^R/HU_{K_U}}=\int_{\tau _{K_V}^R}\Gamma \left( -y_{K_V}^R\right)
\left( \frac{HV_{K_V}^R}{HU_{K_U}}\right) ^{y_{K_V}^R},  \label{dist1}
\end{equation}
where $\int_{\tau _{K_V}^R}$ is a short notation for $\int_{-\infty }^{+\infty }%
\frac{d(\text{Im }y_{K_V}^R)}{2\pi }$, with Re $y_{K_V}^R$ fixed at $\tau _{K_V}^R<0$.
However, for the imaginary part one cannot apply anymore the same identity.
It nevertheless remains true in the sense of distributions. More precisely,
we have for $HV_{K_V}^R/HU_{K_U}>0$ and $-1<\tau _{K_V}^I<0$ (see \cite
{rivasseau7}) 
\begin{equation}
e^{-HV_{K_V}^I/HU_{K_U}}=\int_{\tau _{K_V}^I}\Gamma \left( -y_{K_V}^I\right)
\left( \frac{i\,HV_{K_V}^I}{HU_{K_U}}\right) ^{y_{K_V}^I},  \label{dist2}
\end{equation}
which introduces another set of Mellin parameters. The distributional sense
of the formula above is a major difference with respect to the commutative
case.

For the polynomial $HU$ one can use the formula \cite{rivasseau7} 
\begin{equation}
\Gamma \left( \sum_{K_V}y_{K_V}+\frac d2\right) \left( HU_{K_U}\right)
^{-\sum_{K_V}\left( y_{K_V}^R+y_{K_V}^I\right) -d/2}=\int_\sigma
\prod_{K_U}\Gamma (-x_{K_U})HU_{K_U}^{x_{K_U}}.  \label{dist3}
\end{equation}

As in the commutative case, we now insert the distribution formulas (\ref
{dist1}), (\ref{dist2}) and (\ref{dist3}) into the general form of the
Feynman amplitude. This gives 
\begin{eqnarray}
\mathcal{A}_G &=&\mathcal{K}_G\int_\Delta \frac{\prod_{K_U}a_{K_U}^{x_{K_U}}%
\Gamma (-x_{K_U})}{\Gamma \left( -\sum_{K_U}x_{K_U}\right) }\left(
\prod_{K_V}\left( s_{K_V}^R\right) ^{y_{K_V}^R}\Gamma \left(
-y_{K_V}^R\right) \right)   \nonumber \\
&&\times \left( \prod_{K_V}\left( s_{K_V}^I\right) ^{y_{K_V}^I}\Gamma \left(
-y_{K_V}^I\right) \right) \int_0^1\prod_{\ell =1}^Ldt_\ell \,(1-t_\ell
^2)^{d/2-1}t_\ell ^{\phi _\ell -1},  \label{form34}
\end{eqnarray}
where 
\begin{equation}
\phi _\ell \equiv \sum_{K_U}u_{\ell K_U}x_{K_U}+\sum_{K_V}\left( v_{\ell
K_V}^Ry_{K_V}^R+v_{\ell K_V}^Iy_{K_V}^I\right) +1.
\label{fi}
\end{equation}
Here $\int_\Delta $ means integration over the variables $\frac{\text{Im }%
x_{K_U}}{2\pi i}$, $\frac{\text{Im }y_{K_V}^R}{2\pi i}$ and $\frac{\text{Im }%
y_{K_V}^I}{2\pi i}$, where $\Delta $ is the convex domain 
\begin{equation}
\Delta =\left\{ \sigma ,\tau ^R,\tau ^I\left| 
\begin{array}{l}
\sigma _{K_U}<0;\,\tau _{K_V}^R<0;\,-1<\tau _{K_V}^I<0; \\ 
\sum_{K_U}x_{K_U}+\sum_{K_V}\left( y_{K_V}^R+y_{K_V}^I\right) =-d/2; \\ 
\forall \ell \text{, Re }\phi _\ell =\sum_{K_U}u_{\ell K_U}x_{K_U} \\ 
+\sum_{K_V}\left( v_{\ell K_V}^Ry_{K_V}^R+v_{\ell K_V}^Iy_{K_V}^I\right) +1>0
\end{array}
\right. \right\} 
\label{Delta}
\end{equation}
and $\sigma $, $\tau ^R$ and $\tau ^I$ stand for Re $x_{K_U}$, Re $y_{K_V}^R$
and Re $y_{K_V}^I$. The $t_\ell $ integrations in (\ref{form34}) may be
performed using the representation for the beta function 
\begin{equation}
\int_0^1dt_\ell \,(1-t_\ell ^2)^{d/2-1}t_\ell ^{\phi _\ell -1}=\frac
12B\left( \frac{\phi _\ell }2,\frac d2\right) =\frac{\Gamma \left( \frac{%
\phi _\ell }2\right) \Gamma \left( \frac d2\right) }{2\Gamma \left( \frac{%
\phi _\ell +d}2\right) }.
\end{equation}
The representation is convergent for $0<$ Re $d<2$. Therefore, we can claim
that any Feynman amplitude of a $\phi ^{\star 4}$ diagram is analytic at
least in the strip $0<$ Re $d<2$, where it admits the following CM
representation \cite{rivasseau7} 
\begin{eqnarray}
\mathcal{A}_G &=&\mathcal{K}_G\int_\Delta \frac{\prod_{K_U}a_{K_U}^{x_{K_U}}%
\Gamma (-x_{K_U})}{\Gamma \left( -\sum_{K_U}x_{K_U}\right) }\left(
\prod_{K_V}\left( s_{K_V}^R\right) ^{y_{K_V}^R}\Gamma \left(
-y_{K_V}^R\right) \right)   \nonumber \\
&&\times \left( \prod_{K_V}\left( s_{K_V}^I\right) ^{y_{K_V}^I}\Gamma \left(
-y_{K_V}^I\right) \right) \left( \prod_{\ell =1}^L\frac{\Gamma \left( \frac{%
\phi _\ell }2\right) \Gamma \left( \frac d2\right) }{2\Gamma \left( \frac{%
\phi _\ell +d}2\right) }\right) ,
\end{eqnarray}
which holds as a tempered distribution of the external invariants.

We have thus obtained the complete Mellin representation of Feynman
amplitudes for a noncommutative quantum field theory. The beta functions,
which result from the $t_\ell $-integrations, lead to the appearance of
gamma functions that were not present in the commutative case. We will comment 
about this in the next section. 

\section{Asymptotic expansions for convergent amplitudes}

A general asymptotic regime is defined by scaling the invariants $s_{K_V}^R$, 
 $s_{K_V}^I$ and $a_{K_U}$,  
\begin{eqnarray}
s_{K_V}^R &\rightarrow &\lambda ^{b_{K_V}}s_{K_V}^R  \nonumber \\
s_{K_V}^I &\rightarrow &\lambda ^{c_{K_V}}s_{K_V}^I  \nonumber \\
a_{K_U} &\rightarrow &\lambda ^{d_{K_U}}a_{K_U},
\label{scaling}
\end{eqnarray}
where $b_{K_V}$, $c_{K_V}$ and $d_{K_U}$ may have positive, negative or null values,
and letting $\lambda $ go to infinity. We then obtain under these scalings 
\begin{eqnarray}
\mathcal{A}_{\mathcal{G}}(\lambda ) &=&\mathcal{K}_G\int_\Delta \frac{%
\prod_{K_U}a_{K_U}^{x_{K_U}}\Gamma (-x_{K_U})}{\Gamma \left(
-\sum_{K_U}x_{K_U}\right) }\left( \prod_{K_V}\left( s_{K_V}^{\mathcal{R}%
}\right) ^{y_{K_V}^{\mathcal{R}}}\Gamma \left( -y_{K_V}^{\mathcal{R}}\right)
\right)  \nonumber \\
&&\times \left( \prod_{K_V}\left( s_{K_V}^{\mathcal{I}}\right) ^{y_{K_V}^{%
\mathcal{I}}}\Gamma \left( -y_{K_V}^{\mathcal{I}}\right) \right) \left(
\prod_{\ell =1}^L\frac{\Gamma \left( \frac{\phi _\ell }2\right) \Gamma
\left( \frac d2\right) }{2\Gamma \left( \frac{\phi _\ell +d}2\right) }%
\right) \,\lambda ^\psi ,
\label{asympt1}
\end{eqnarray}
where the exponent of $\lambda $ is a linear function of the Mellin
variables: 
\begin{equation}
\psi =\sum_{K_{V},K_U}\left( b_{K_V}y_{K_V}^R+c_{K_V}y_{K_V}^I+d_{K_U}x_{K_U}\right) .
\end{equation}
Notice that the factor  
 $\left[ \prod_{\ell =1}^L\Gamma \left( \frac d2\right)2\Gamma \left( 
\frac{\phi _\ell +d}2\right) \right] ^{-1}$ in the integrand of eq.~(\ref{asympt1}) does
 not affect the meromorphic
structure of the amplitude (\ref{asympt1}). Moreover, for strictly positive dimensions 
$d>0$ and $\phi _\ell \in \Delta $, this factor  also does not introduce $zeroes$ in the integrand. 

From the above expressions, we can show that the proof of the theorem given
in \cite{decalan1} can be extended for the noncommutative case. To do this,
let us rewrite the above expression for $\mathcal{A}_G(\lambda )$ in a
convenient way. Let us denote collectively the variables $\left\{
x_{K_U},y_{K_V}\right\} $ as $\left\{ z_K\right\} $, whereas the arguments
of the gamma functions leading to singularities, $-x_{K_U}$, $-y_{K_V}^R$, $%
-y_{K_V}^I$, and $\frac{\phi _\ell }2$ will be renamed $\psi _\nu (z_K)$.
The convex domain $\Delta $ can then be rewritten simply as 
\begin{equation}
\Delta =\left\{ z_K\text{ such that Re }\psi _\nu (z_K)>0\text{, for all }%
\nu \right\}, 
\end{equation}
Let us define the set of quantities $\{s_\nu \}$ such that it includes the
quantity $a_{K_U}$, which are functions of the objects $\Theta $ and $\Omega 
$ having no correspondents in ordinary commutative field theory, 
\begin{equation}
s_\nu =\left\{ 
\begin{array}{ccc}
a_{K_U} &  & \text{if }z_K=x_{K_U} \\ 
s_{K_V}^R &  & \text{if }z_K=y_{K_V}^R \\ 
s_{K_V}^I &  & \text{if }z_K=y_{K_V}^I \\ 
1 &  & \text{if }z_K=\frac{\phi _\ell }2.
\end{array}
\right. 
\end{equation}
Also, we introduce the factors $f_{\nu }$ (in general functions of the variables 
$x_{K_U}$ and $y_{K_V}$) such that
\begin{equation}
f_{\nu }=\left\{ 
\begin{array}{lll}
\Gamma \left( \frac d2\right)\left[ \prod_{\ell =1}^L2\Gamma \left( 
\frac{\phi _\ell +d}{2}\right) \right] ^{-1} &  & \text{if }\psi_{\nu}=\phi _{\ell }/2  \\ 
1 &  & \text{otherwise}.
\end{array}
\right. 
\end{equation}
Therefore the expression for $\mathcal{A}_G(\lambda )$ in (\ref{asympt1}) 
can be simplified to 
\begin{equation}
\mathcal{A}_G(\lambda )=\mathcal{K}_G\int_\Delta \lambda ^\psi \prod_{\nu }f_{\nu}\,s_\nu
^{-\psi _\nu }\Gamma (\psi _\nu )\frac 1{\Gamma \left(
-\sum_{K_U}x_{K_U}\right) }.  \label{asympt2}
\end{equation}

Eq.~(\ref{asympt2}) has exactly the same singularity structure as (\ref
{lambfunc}), the factors $f_{\nu }$ only modify the residues at the poles. 
Thus we can translate to the
present situation all the steps of the proof of the asymptotics theorem of
Ref.~\cite{decalan1}, since it relies entirely on displacements of the integration contours crossing the singularities of the gamma functions $\Gamma (\psi _{\nu })$ [$\Gamma (\zeta _\mu )$, in the commutative counterpart of eq.~(\ref{lambfunc})]. For completeness, this demonstration is given in the Appendix. Thus the  result of
Ref.~\cite{decalan1} remains valid {\it mutatis mutandis} for the the vulcanized $\phi ^{\star 4}$ theory and we are allowed to state the following theorem: 

\begin{theorem}\label{cmasympt}
Let us consider a ribbon diagram $G$ of the vulcanized $\phi ^{\star 4}$ theory,  
and its related amplitude $\mathcal{A}_G(\lambda )$ 
 under the general scaling of its invariants,
\begin{eqnarray}
s_{K_V}^R &\rightarrow &\lambda ^{b_{K_V}}s_{K_V}^R  \nonumber \\
s_{K_V}^I &\rightarrow &\lambda ^{c_{K_V}}s_{K_V}^I  \nonumber \\
a_{K_U} &\rightarrow &\lambda ^{d_{K_U}}a_{K_U},
\label{scaling}
\end{eqnarray}
where $b_{K_V}$, $c_{K_V}$ and $d_{K_U}$ may have positive, negative or null values,
and as $\lambda \rightarrow \infty $. Then 
there exists an asymptotic expansion of $\mathcal{A}_G(\lambda )$ of the
form 
\begin{equation}
\mathcal{A}_G(\lambda )=\sum_{p=p_{\text{max}}}^{-\infty }\sum_{q=0}^{q_{%
\text{max}}(p)}\mathcal{A}_{pq}(s_{K_V}^R,s_{K_V}^I,a_{K_U})\,\lambda ^p\ln
^q\lambda ,  \label{asympt}
\end{equation}
where $p$ runs over the rational values of a decreasing arithmetic
progression, with $p_{\text{max}}$ as a `leading power', and $q$, for a
given $p$, runs over a finite number of nonnegative integer values.
\end{theorem}
The coefficients $\mathcal{A}_{pq}(s_{K_V}^R,s_{K_V}^I,a_{K_U})$ of the expansion in (\ref{asympt}) are functions only of the invariants associated to the hyperbolic polynomials. Notice, in particular, that the invariants $a_{K_U}$ contain the noncommutative entities $\Theta $ and $\Omega $ encoded.

\section{Renormalized amplitudes}

Let us now consider the complete Mellin representation of divergent
amplitudes. The analysis follows the steps taken in \cite{decalan1,decalan2}
for the commutative case, and the recent results of Ref.~\cite{gurau} for
noncommutative theories. We have to go back to the amplitude given in eq.~(%
\ref{param}), for the occurrence of ultraviolet divergences in an expression
such as this one prevents the interchange of the integral over the $t_\ell $
variables with the one over the domain $\Delta $. It means that the $t_\ell $%
-integral cannot be performed and a renormalization prescription is
therefore required. For this, we use the method of subtracting the first
terms of a generalized Taylor expansion corresponding to the infinities of
the divergent subdiagrams \cite{bergere1,bergere2}, as adapted to the $%
t_\ell $ integrations \cite{gurau}. Each $t_\ell $ variable belonging to a
divergent subdiagram $S$ is scaled by a parameter $\rho ^2$, $t_{\ell \in
S}\rightarrow \rho ^2t_{\ell \in S}$, and the integral in eq.~(\ref{param})
becomes a function of $\rho $, which we call $g(\rho )$. Next, following the
steps of \cite{bergere1}, we define the generalized Taylor operator of order 
$n$, 
\begin{equation}
\tau ^n\left[ \rho ^\nu g(\rho )\right] =\rho ^\nu T_\rho ^{n-E[\nu ]}\left[
g(\rho )\right] ,
\end{equation}
with $E[\nu ]$ being the smallest integer greater than $\nu $, and $T_\rho
^q\left[ g(\rho )\right] $ being the (truncated) usual Taylor operator over
a function $g(\rho )$, 
\begin{equation}
T_\rho ^q\left[ g(\rho )\right] =\sum_{k=0}^q\frac{\rho ^k}{k!}g^{(k)}(0),
\label{taylor}
\end{equation}
which makes sense only for $q\geq 0$. The generalized Taylor operator acts
on the $t_\ell $ integrand, so that for each primitively divergent
subdiagram $S$ of $G$ one associates a subtraction operator $\tau _S^{-2l_S}$%
, where $l_S$ is the number of lines in the subdiagram $S$. The $\tau _S$
operator is equivalent to the introduction of counterterms in the theory, in
that it is defined in order to suppress the ultraviolet divergent terms from
the integrand; the $t_\ell $ variables associated to the subdiagram $S$ are
first scaled by the parameter $\rho $, and the first few terms of the
generalized Taylor expansion in $\rho $ are kept in $\mathcal{\tau }_S$. In
fact, this corresponds to the Taylor operator in eq.~(\ref{taylor}),
truncated at the order $q$, which is the superficial degree of convergence
(the negative of the superficial degree of divergence) of the subdiagram $S$%
, $q=dL_S-2l_S$, $L_S$ being the number of loops of $S$. At the end of the
that computation, one takes $\rho =1$.

Now, a crucial point, argued in Ref. \cite{gurau}, is that since one is
interested in the region of ultraviolet divergences, the factor $\prod_\ell
dt_\ell \,(1-t_\ell ^2)^{d/2-1}$ in eq.~(\ref{param}) can be bounded in such
a way that it cannot contribute to divergences and so it is included in the
integration mesure. This factor plays exactly the same r\^{o}le of the
integration measure $\prod_\ell d\alpha _\ell \exp \left( -\sum_\ell m_\ell
\alpha _\ell \right) $ in the massive commutative case. Thus the action of
the generalized Taylor operator on the integrand is 
\begin{equation}
\tau _S^{-2l_S}\left( \frac{e^{-HV_G/HU_G}}{HU_G^{d/2}}\right) =\left[ \tau
^{-2l_S}\left( \left. \frac{e^{-HV_G/HU_G}}{HU_G^{d/2}}\right|
_{t_S\rightarrow \rho ^2t_S}\right) \right] _{\rho =1}.
\end{equation}

The renormalized amplitude is defined by introducing the operator $\mathcal{R%
}$ \cite{bergere1}, 
\begin{equation}
\mathcal{R}=\prod_S\left( 1-\tau _S^{-2l_S}\right) ,  \label{R}
\end{equation}
which satisfies the identity \cite{gurau} 
\begin{equation}
\mathcal{R}=1+\,\sum_{\mathcal{F}}\prod_{S\in \mathcal{F}}(-\tau
_S^{-2l_S})=\prod_S\left( 1-\tau _S^{-2l_S}\right) ,  \label{R1}
\end{equation}
where $\mathcal{F}$ is the set of all nonempty forests of primitively
divergent subdiagrams. Then the renormalized amplitude is 
\begin{equation}
\mathcal{A}_G^{\text{ren}}(\{x_e\},p_{\bar{V}})=K^{\prime
}\int_0^1\prod_\ell dt_\ell \,(1-t_\ell ^2)^{d/2-1}\mathcal{R}\left\{ \frac{%
e^{-HV_{G,\bar{V}}(t,x_e,p_{\bar{V}})/HU_{G,\bar{V}}(t)}}{\left[ HU_{G,\bar{V%
}}(t)\right] ^{d/2}}\right\} .
\end{equation}

Now, within the context of the complete Mellin representation, we have,
from (\ref{form34}), 
\begin{eqnarray}
\mathcal{A}_G^{\text{ren}} &=&\mathcal{K}_G\int_\Delta \frac{\prod_{K_U}a_{K_U}^{x_{K_U}}%
\Gamma (-x_{K_U})}{\Gamma \left( -\sum_{K_U}x_{K_U}\right) }\left(
\prod_{K_V}\left( s_{K_V}^R\right) ^{y_{K_V}^R}\Gamma \left(
-y_{K_V}^R\right) \right)   \nonumber \\
&&\times \left( \prod_{K_V}\left( s_{K_V}^I\right) ^{y_{K_V}^I}\Gamma \left(
-y_{K_V}^I\right) \right) \int_0^1\prod_{\ell =1}^Ldt_\ell \,(1-t_\ell
^2)^{d/2-1}\mathcal{R}\left[ t_\ell ^{\phi _\ell -1}\right] . \\ \label{ampR}
\end{eqnarray}
This is the analogous of the starting point of the analysis of \cite
{decalan2} on renormalized amplitudes in the complete Mellin representation.
We see that the renormalization operator $\mathcal{R}$ acts on the $t_\ell $%
-variables, \textit{exactly in the same way} as it acts on the $\alpha _l$%
-variables in the commutative situation of Refs. \cite{decalan1,decalan2},
the only difference (which does not affect the validity of the theorems in 
\cite{decalan2}) being in the integration measure over the $t_\ell $%
-variables.

Therefore, the theorems stated in \cite{decalan2} remain valid in the
vulcanized noncommutative case. Then we can follow the same steps as in
Ref.~\cite{decalan2}, that is, we define cells $C$, such that 
\begin{equation}
\sup_C\text{Re}(\phi _\ell )>0\,,\forall \ell ;\qquad \inf_S\left(
\inf_C\sum_{\ell \in S}\text{Re}(\phi _\ell )\right) \leq 0.  \label{C}
\end{equation}
The effect of the $\mathcal{R}$ operator in eq.~(\ref{ampR}) is to split the
factor $\mathcal{R}\left( t_\ell ^{\phi _\ell -1}\right) $ into a set of
terms $\left\{ \mu _Ct_\ell ^{\phi _\ell -1},\phi _\ell \in C\right\} $,
where $\mu _C$ are numerical coefficients. This allows the $t_\ell $
integral to be evaluated just as in the convergent case. The renormalized
amplitude in the CM representation is then given by 
\begin{equation}
\mathcal{A}_G^{\text{ren}}=\mathcal{K}_G\sum_C\mu _C\int_{\Delta _C}\mathcal{I}%
_C(x_{K_U},y_{K_V}^R,y_{K_V}^I),
\end{equation}
where we have defined the integrands 
\begin{eqnarray}
\mathcal{I}_C &=&\frac{\prod_{K_U}a_{K_U}^{x_{K_U}}\Gamma (-x_{K_U})}{\Gamma
\left( -\sum_{K_U}x_{K_U}\right) }\left( \prod_{K_V}\left( s_{K_V}^R\right)
^{y_{K_V}^R}\Gamma \left( -y_{K_V}^R\right) \right)   \nonumber \\
&&\times \left( \prod_{K_V}\left( s_{K_V}^I\right) ^{y_{K_V}^I}\Gamma \left(
-y_{K_V}^I\right) \right) \left( \prod_{\ell =1}^L\Gamma \left( \frac{\phi
_\ell }2\right) \right).  \label{CMfinal}
\end{eqnarray}
We now have a set of integration domains given by 
\begin{equation}
\Delta _C=\left\{ \sigma ,\tau ^R,\tau ^I\in C\left| 
\begin{array}{l}
\sigma _{K_U}<0;\,\tau _{K_V}^R<0;\,-1<\tau _{K_V}^I<0; \\ 
\sum_{K_U}x_{K_U}+\sum_{K_V}\left( y_{K_V}^R+y_{K_V}^I\right) =-d/2; \\ 
\forall \ell \text{, Re }\phi _\ell =\sum_{K_U}u_{\ell K_U}x_{K_U} \\ 
+\sum_{K_V}\left( v_{\ell K_V}^Ry_{K_V}^R+v_{\ell K_V}^Iy_{K_V}^I\right) +1>0
\end{array}
\right. \right\} ,  \label{deltaC}
\end{equation}
instead of the single one ($\Delta $) of the convergent amplitude. As in the
commutative case, we see that the renormalization procedure only changes the
relevant integration domains in the Mellin variables. The structure of the
integrands $\mathcal{I}_C$ in the cells $C$ remains exactly of the same form
as for convergent amplitudes. This then implies that we can apply in each
cell the machinery used in the previous section and we can state the following theorem:  

\begin{theorem}\label{cmasympt1}
Under
the scaling of eq.~(\ref{scaling}), each integral $\int_{\Delta _C}\mathcal{I%
}_C(x_{K_U},y_{K_V}^R,y_{K_V}^I)$ has an asymptotic expansion of the same form 
of the the one of eq.~(\ref{asympt}); therefore the amplitude $\mathcal{A}_G^{\text{ren}}(\lambda )$ 
has an asymptotic expansion of the form  
\begin{equation}
\mathcal{A}_G^{\text{ren}}(\lambda )=\sum_C\mu _C\mathcal{I}_C(\lambda ),
\end{equation}
with 
\begin{equation}
\mathcal{I}_C(\lambda )=\sum_{p=p_{\text{max}}^C}^{-\infty }\sum_{q=0}^{q_{%
\text{max}}^C(p)}\mathcal{A}_{pq}^C(s_{K_V}^R,s_{K_V}^I,a_{K_U})\,\lambda ^p\ln
^q\lambda  
\end{equation}
and where in each cell $C$, $p$ runs over the rational values of a decreasing arithmetic
progression, with $p^{C}_{\text{max}}$ as a `leading power', and $q$, for a
given $p$, runs over a finite number of nonnegative integer values.
\end{theorem} 
As in the convergent case, the coefficients $\mathcal{A}_{pq}^{C}(s_{K_V}^R,s_{K_V}^I,a_{K_U})$ of the expansion in $\mathcal{I}_C(\lambda )$ are functions only of the invariants associated to the hyperbolic polynomials. 

\section{Conclusions}

We have shown in this paper that all the steps in the proofs of the theorems 
in Refs.~\cite{decalan1,decalan2} may be reproduced in the context of the vulcanized
noncommutative scalar $\phi ^{*4}$ model. In particular, the proof of the 
existence of asymptotic expansions for Feynman amplitudes in commutative
field theories done in \cite{decalan1} may be transposed to the present situation,  
for both convergent and renormalized amplitudes. The resulting theorems 
take into account the influence of the specificities of the noncommutative
generalization of the theory in the details of the proof. In particular, it
was crucial to observe that the parameters $a_{K_U}$, within which the
noncommutative entities $\Theta $ and $\Omega $ are encoded, and are of
course inexistent in the commutative case, may be defined as part of the
`invariants' $s_\nu $ and therefore are related to the meromorphic structure
of the amplitude and its asymptotic behaviour can be studied. Another difference with respect to the commutative case is
the fact that the next set of invariants $s_\nu $, $s_{K_V}$, have real and imaginary
parts ($s_{K_V}^R$ and $s_{K_V}^I$), and they contribute separately. Also, as the field we are considering
is massless, the $s_\nu $ related to the functions $\phi _\ell $ are
trivial. In principle, the explicit calculation of the coefficients of the expansions, in both theorems \ref{cmasympt} and \ref{cmasympt1}, is
possible but, for a general amplitude, is an extremely hard task. Nevertheless, those corresponding to the leading terms can be evaluated (see Appendix) along the same
lines as in the commutative case in Ref.~\cite{decalan1}.

\section{Appendix: Proof of the theorem 4.1}
\label{dem}
In this Appendix we perform a ``translation" for the noncommutative theory, of the proof of the 
asymptotics theorem of Ref.~\cite{decalan1}. In eq.~(\ref{asympt2}), when $\psi
_\nu \in \Delta $, the integral is absolutely convergent, so we have a
first bound: 
\begin{equation}
\mathcal{A}_{\mathcal{G}}(\lambda )<\text{ const. }\lambda ^{p_{\text{max}%
}+\epsilon }\,;\;\;\;\;p_{\text{max}}=\inf_{\Delta }\left( \text{Re }\psi(z_K)\right),
\end{equation}
where $\epsilon $ is an arbitrary small number.  
Therefore, the function $\psi(z_K)-p_{\text{max}}$ is positive in $\Delta $,
and reaches zero on its boundary. It then ensues that there exist
nonnegative coefficients $d_\nu $ such that $\psi(z_K)-p_{\text{max}}=\sum_\nu d_\nu \psi _\nu$, which implies 
\begin{equation}
\frac 1{\prod_\nu \psi _\nu }\equiv \frac 1{\psi(z_K)-p_{\text{max}%
}}\sum_\nu \frac{d_\nu }{\prod_{\nu ^{\prime }\neq \nu }\psi _{\nu ^{\prime
}}}.
\end{equation}
For a given $\nu $, if the subset $\left\{ \psi _{\nu ^{\prime }},\text{ }%
\nu ^{\prime }\neq \nu \right\} $ still generates $\psi(z_K)-p_{\text{max}}$%
, we can repeat the procedure, which is iterated until we obtain 
\begin{equation}
\frac 1{\prod_\nu \psi _\nu }\equiv \sum_E\frac{d_E}{\left( \psi(z_K)-p_{%
\text{max}}\right) ^{q_E+1}}\frac 1{\prod_{\nu \in E}\psi _\nu }. 
\end{equation}
For each $E\subset \{\nu \}$, $\psi(z_K)-p_{\text{max}}$ does not belong to the convex
domain defined by the subset $\left\{ \psi _\nu \geq 0,\text{ }\nu \in
E\right\} $ and it becomes negative somewhere in 
\begin{equation}
\Delta _E=\left\{ z_K\text{ such that }\psi _\nu +\theta _{\nu E}>0,\text{
for all }\nu \right\}\,;\;\;\;\;\theta _{\nu E}=\left\{ 
\begin{array}{ccc}
0 &  & \text{if }\nu \in E \\ 
1 &  & \text{otherwise}.
\end{array}
\right. 
\end{equation}
Therefore, the amplitude $\mathcal{A}_{\mathcal{G}}(\lambda )$ in eq~(\ref{asympt2})
becomes,
\begin{equation}
\mathcal{A}_{\mathcal{G}}(\lambda )=\sum_Ed_E\int_{\Delta _E;\,\text{Re }%
(\psi(z_K)-p_{\text{max}})>0}\frac{\lambda ^\psi M_E(z_K)}{\left( \psi(z_K)-p_{\text{%
max}}\right) ^{q_E+1}},
\end{equation}
where we have defined the function 
\begin{equation}
M_E(z)=\frac{\prod_\nu f_{\nu }s_\nu ^{-\psi _\nu }\Gamma (\psi _\nu +\theta _{\nu
E}) }{\Gamma \left( -\sum_{K_U}x_{K_U}\right) },
\end{equation}
which is analytical in $\Delta _E$.

The integration path can be moved up to a point where $\psi(z_K)-p_{\text{max%
}}<0$, and applying Cauchy's integral formula we obtain 
\begin{equation}
\mathcal{A}_{\mathcal{G}}(\lambda )=\sum_Ed_E\,\left\{ \lambda ^{p_{\text{max%
}}}\sum_{q=0}^{q_E}\mathcal{A}_{p_{\text{max}}q}^E\,\ln ^q\lambda
+\int_{\Delta ^{'}_E}\frac{\lambda ^\psi M_E(z_K)
}{\left( \psi(z_K)-p_{\text{max}}\right) ^{q_E+1}}\right\} ,
\label{aftercauchy}
\end{equation}
where in $\Delta ^{'}_E:\,\text{Re }(\psi(z_K)-p_{\text{max}})<0$ and 
\begin{equation}
\mathcal{A}_{p_{\text{max}}q}^E=\frac 1{q!\,\left( q_E-q\right)
!}\int_{\Delta _E;\,\psi(z_K)-p_{\text{max}}=0}\nabla ^{q_E-q}M_E(z_K),
\end{equation}
with $\nabla $ being the differential operator along any direction crossing
the plane $\psi=p_{\text{max}}$. The integral in the second term of (\ref
{aftercauchy}) is bounded, being less than a constant times $\lambda ^{p_{%
\text{max}}-b_E+\epsilon }$, where $p_{\text{max}}-b_E=$ Inf$_{\Delta
_E}\left( \text{Re }\psi(z_K)\right) $, in which $b_E$ is a strictly
positive rational number.

The remaining gamma-function singularities are treated in a similar fashion,
by applying the identity 
\begin{equation}
\Gamma (\psi _\nu +\theta _{\nu E})=\frac 1{\psi _\nu +\theta _{\nu
E}}\Gamma (\psi _\nu +\theta _{\nu E}+1)
\end{equation}
in the second term of (\ref{aftercauchy}), leading to an analogous term with 
$\lambda ^{p_{\text{max}}-a_E}$, and another integral in the next
analiticity strip, and so forth. In this way, a complete asymptotic
expansion is produced.

\end{document}